\documentclass[useAMS,usenatbib]{mn2e}
\usepackage[pdftex]{graphicx} 
\usepackage[hidelinks]{hyperref}
\usepackage{xcolor} 
\hypersetup{colorlinks,linkcolor={red!50!black}, citecolor={blue!50!black},urlcolor={blue!80!black} } 

\title[An observational proxy of halo assembly time and 
its correlation with galaxy properties]{An observational proxy of halo assembly time and 
its correlation with galaxy properties}

\author[S.H. Lim et al.]{S.H. Lim$^{1}$\thanks{E-mail:
slim@astro.umass.edu}, H.J. Mo$^{1}$, Huiyuan Wang$^{2}$ and Xiaohu Yang$^{3,4}$\\
$^{1}$Department of Astronomy, University of Massachusetts, Amherst MA 01003-9305, USA\\
$^{2}$Key Laboratory for Research in Galaxies and Cosmology, University of Science and 
Technology of China, Hefei, Anhui 230026, China\\
$^{3}$Center for Astronomy and Astrophysics, Shanghai Jiao Tong University, Shanghai 200240, China\\
$^{4}$IFSA Collaborative Innovation Center, Shanghai Jiao Tong University, Shanghai 200240, China}

\begin{document} 

\date{Accepted ........ Received .......; in original form ......}

\pagerange{\pageref{firstpage}--\pageref{lastpage}}

\pubyear{2015}

\maketitle

\label{firstpage}

\begin{abstract} 
We show that  the ratio between the stellar mass of central galaxy 
and the mass of its host halo, $f_c \equiv M_{*,c}/M_{\rm h}$, can be used as 
an observable proxy of halo assembly time, in that galaxy groups 
with higher $f_c$ assembled their masses earlier. Using SDSS groups of Yang et al.,
we study how $f_c$ correlates with galaxy properties such as 
color, star formation rate, metallicity, bulge to disk ratio, 
and size. Central galaxies of a given stellar mass in groups with $f_c>0.02$ 
tend to be redder in color, more quenched in star formation, 
smaller in size, and more bulge dominated, as $f_c$ increases. The trends 
in color and star formation appear to reverse at $f_c<0.02$, reflecting a 
down-sizing effect that galaxies in massive halos formed their stars 
earlier although the host halos themselves assembled later (lower $f_c$). No such 
reversal is seen in the size of elliptical galaxies, suggesting that their 
assembly follows halo growth more closely than their star formation. 
Satellite galaxies of a given stellar mass in groups of a given halo 
mass tend to be redder in color, more quenched in star formation 
and smaller in size as $f_c$ increases. For a given stellar mass, 
satellites also tend to be smaller than centrals.  The trends are 
stronger for lower mass groups. For groups more massive than 
$\sim 10^{13}{\rm M}_\odot$, a weak reversed trend is seen in color 
and star formation. The observed trends in star formation are 
qualitatively reproduced by an empirical model based on halo age 
abundance matching, but not by a semi-analytical model tested here. 
\end{abstract} 


\section{Introduction}
\label{sec_intro}

In the current standard $\Lambda$CDM model, dark matter halos form through 
gravitational instability - induced hierarchical clustering, and galaxies are believed 
to form at the centers of dark matter halos through cooling and condensation 
of baryonic gas \citep[e.g.][for a review]{mo10}.
The formation and evolution of galaxies are, therefore, 
expected to be closely linked to the assembly history of their host halos. 
There have been continuous efforts to establish the connections between galaxies of 
different properties and dark matter halos using empirical models, such as 
halo occupation distribution (HOD) \citep[e.g.][]{jing98,peacock00, seljak00, scoccimarro01, berlind02,
zheng07, leauthaud12, watson12},  conditional luminosity function
(CLF) \citep{yang03, vandenbosch07}, and halo abundance matching (HAM) 
\citep[][]{MoMaoWhite99, kravtsov04, vale04, vale06, conroy09, guo10, neistein10, watson12, kravtsov13}. 
The CLF and HOD models assign galaxies into dark matter halos predicted by a given 
cosmology, so that the predicted galaxy population matches the observed 
luminosity (stellar mass) functions and spatial clustering properties of 
galaxies. The HAM approach, on the other hand, populates galaxies into halos and sub-halos, 
assuming that there is a roughly monotonic correspondence between the ranking 
orders of the luminosities (or stellar masses) of galaxies and those of the masses of 
dark matter halos.  

Most of the studies based on these approaches have so far focused on 
using the mass of halos to link galaxies with halos, thus implicitly assuming that 
galaxy properties are determined by halo mass alone. In reality, however, 
other properties of halos, such as assembly history, spin, and shape, may also 
play an important role in galaxy formation and evolution. These halo properties, therefore, 
should also be used in understanding the relationships between galaxies and halos. 

In this paper, we investigate how the properties of galaxies of a given 
stellar mass are correlated with the assembly time of their host halos. 
To this end, we propose an observational proxy of halo assembly time
motivated by the results of \citet{wang11}. Using high-resolution $N$-body 
simulations, Wang et al.  investigated a large number of halos properties, 
such as formation time, substructure fraction, spin and shape, and their 
correlations among themselves and with large scale environments. 
Most of these halo properties are, unfortunately, not directly observable, 
and so it is difficult to test directly their effects on galaxy formation 
with observational data. One exception is the sub-structure fraction, 
which is defined as $f_s= 1-(M_{\rm main}/M_{\rm h})$, 
where $M_{\rm h}$ is the mass of the halo, and $M_{\rm main}$ is the 
mass of the main sub-halo located at the center of the host halo. 
This quantity is found to be correlated tightly with many other halo 
properties, in particular the formation time, spin and shape.  
More importantly, this quantity may be estimated from observations. 
Indeed, with a well-defined galaxy system, such as a galaxy group 
selected with the halo-based group finder of \citet{yang05}, 
a good proxy of $M_{\rm main}$ is $M_{*,c}$, the stellar mass of 
the central galaxy in a group according to halo-galaxy abundance 
matching,  and $M_{\rm h}$ 
can be estimated from the total stellar mass of the group, as
demonstrated in \citet{yang05, yang07}. Thus, one can use $f_c \equiv
M_{*,c}/M_{\rm h}$ as an observational proxy of the assembly time
of the host halo of the group, and study how galaxy properties change 
with $f_c$. The goal of the present paper is to use this proxy to 
study the correlations between galaxy properties and the assembly time 
of their host groups (halos). 

This paper is organized as follows. In Section \ref{sec_data}, we describe 
the observational galaxy catalogs from which our galaxy and group
samples are selected. In Section \ref{sec_theproxy} we demonstrate how $f_c$ 
can be used as a reliable proxy of halo assembly time. Detailed analyses 
of the correlations between galaxy properties and $f_c$ of their host groups 
are presented in Section \ref{sec_correlation}, and a preliminary comparison 
of our results with models is made in Section \ref{sec_theory}.
Finally, in Section \ref{sec_summary}, we summarize our main conclusions. 

\section[Observational data]{Observational Data}
\label{sec_data}

\begin{figure*}
\hspace*{-0.2cm}
\includegraphics[width=1.02\linewidth]{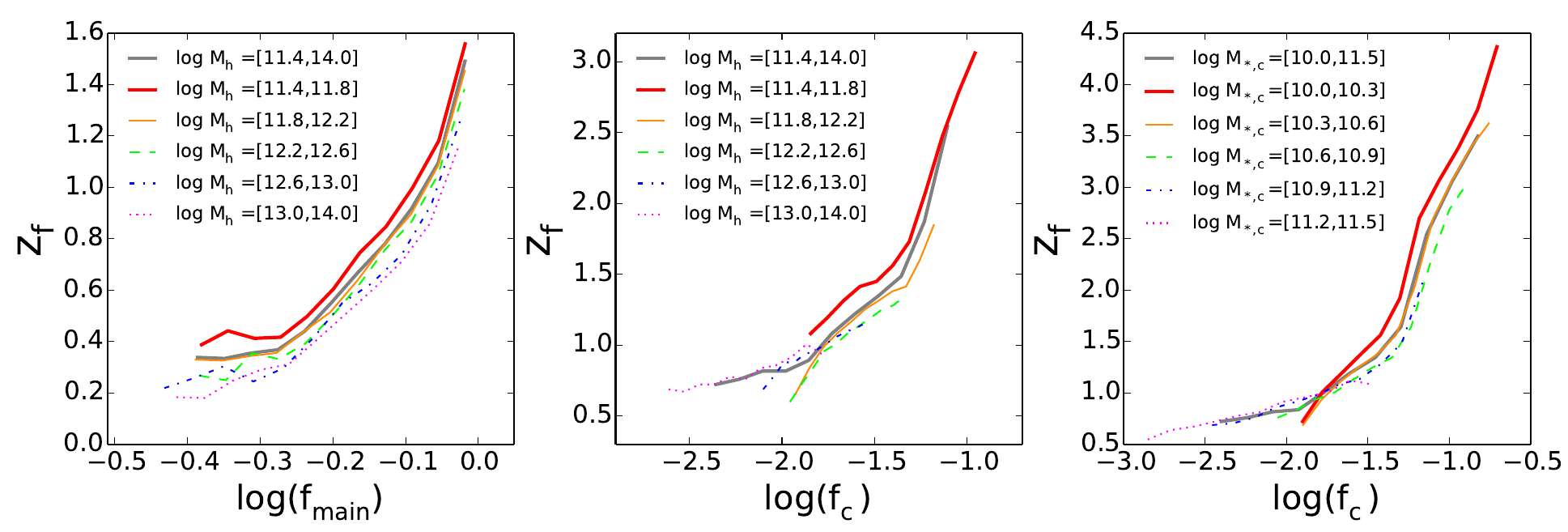}
\caption{
A demonstration how $f_c$ can be used as a proxy of halo assembly time.
{\it Left} : The correlation between half-mass assembly time $z_f$ and
$f_{\rm main}=M_{\rm main}/M_{\rm h}$ (median) obtained from N-body simulations, 
based on data published in  W11, where $M_{\rm main}$ is the mass of the most massive 
sub-halo in each host halo. Results are shown for halos in five mass ranges, as indicated.
For comparison, the result for the total halo sample is shown as the gray line.   
{\it Middle} : The correlation between $z_f$ and $f_c \equiv M_{*,c}/M_{\rm h}$ (median), 
where $M_{*,c}$ is the stellar mass of the central (most massive) galaxy, obtained from
the mock galaxy catalog of \citet{hearin13}, constructed using an age abundance 
matching model combined with halos from the Bolshoi $N$-body simulation. 
Different curves denote different host halo mass bins, as indicated.
The results for the total sample is shown as the gray line.   
{\it Right} : The same as the middle panel, except that 
different curves show different stellar mass bins of central 
galaxies, as indicated. Here again the result for the total sample 
is shown as the gray line for comparison.
}
\label{fig_proxy} 
\end{figure*}

\subsection{SDSS galaxies} 

The galaxy samples used in this paper are obtained from the Sloan 
Digital Sky Survey (SDSS). Specifically, the
galaxy catalog, as described in \citet{wang12} (W12 hereafter) and publicly available 
at \url{http://gax.shao.ac.cn/data/Group.html},  is constructed from the
New York University Value-Added Galaxy Catalogue \citep[NYU-VAGC ;][]{nyu}, 
which is based on SDSS Data Release 7
\citep[SDSS DR7 ;][]{dr7}, but updated with a set of
improvements over the original pipeline. From this catalog, we
select all galaxies in the Main Galaxy Sample with
extinction-corrected apparent $r$-band magnitude brighter
than $17.72$, with redshifts in the range $0.01\leq z \leq0.20$,
and with redshift completeness $C_z > 0.7$. This leaves 639,359
galaxies in total, with a sky coverage of 7,748 $\rmn{deg}^2$. Of
these, 599,301 galaxies have redshifts from the SDSS DR7, 2,450 galaxies
with redshifts from the 2dFGRS \citep{2df},
819 with redshifts from the Korea Institute for
Advanced Study Value-Added Galaxy Catalogue \citep[KIAS-VAGC ; e.g.][]
{kias}, 36,759 galaxies with redshifts from their
nearest neighbors (since they do not have spectroscopic 
redshift measurements due to fiber collisions), and 30 galaxies 
with redshifts from ROSAT X-ray clusters. We exclude galaxies 
with assigned redshifts that have $^{0.1}M_r-5\log h\leq-22.5$ 
to prevent fiber-collided galaxies with real redshifts much lower than 
the nearest neighbors so that their luminosities are vastly over-estimated.    
The catalog also contains, for each galaxy,  the $(g-r)$ and other 
colors,  which are all  $K+E$-corrected to $z=0.1$. 
In the following, this catalog will be referred to as the SDSS DR7 catalog 
to distinguish it from other catalogs we use in our study.

For all galaxies, we adopt stellar masses ($M_*$)  from 
the data release of \citet{brinchmann04}, available at
\url{http://www.mpa-garching.mpg.de/SDSS/DR7/}. 
The data release also provides star formation rates (SFRs),
 and specific star formation rates (sSFRs, defined to be SFR divided by 
 $M_*$).  The SFRs are obtained by fitting the SDSS spectra with 
a spectral synthesis model. Specifically, H$\alpha$ luminosities 
are used  for star forming galaxies  and the D4000 breaks are 
used for galaxies without significant emission lines. Gas phase 
metallicities [for example, oxygen abundance, in terms of log(O/H)] 
are also available for a fraction of the galaxies, as described in \citet{tremonti04}.
In total, about $6\%$ of the galaxies in the SDSS DR7 catalog are
missing in the Brinchmann et al. data release, most of which are
fiber-collided galaxies missing spectra. The number of galaxies 
for which a given quantity is actually available varies from quantity 
to quantity. For example, gas phase metallicity is available only
for emission line galaxies.  

\subsection{Disk-bulge decomposition}

We also make use of the results of \citet{simard11} obtained   
from bulge-disk decompositions of galaxies,  which fit each galaxy 
image with the sum of a pure exponential disk and a de Vaucouleurs 
bulge using GIM2D. The code returns parameters such as 
the total flux, the bulge to total ratio $B/T$, the bulge half-light radius
$R_{50}$ and the disk scale length $R_{\rm disk}$. 
In this paper we use the results  based on the  $r$-band images.  
About $92\%$ of our SDSS DR7 galaxies can be cross identified 
in Simard et al.'s data base.

\subsection{Information from the Galaxy Zoo}

The Galaxy Zoo is a project in which volunteers are asked to 
classify images of over 900,000 SDSS DR7 galaxies into six
morphological categories. The Galaxy Zoo 2 \citep[GZ2 hereafter;][]
{willett13}, the successor of the original Galaxy Zoo, 
continued the spirit of the original project but asking
volunteers much more detailed morphological questions such as
the number of spiral arms, tightness of the arms, etc. 
To enable such detailed questions, GZ2 uses a subsample of the
brightest 25$\%$ of the resolved galaxies in the SDSS North
Galactic Cap region within the redshift range of $0.0005<z<0.25$
along with a few more selection criteria \citep[see][]{willett13}. 
This leaves a grand total of 245,609 SDSS DR7 galaxies.

The SDSS metadata for GZ2 (available at
\url{http://data.galaxyzoo.org/}) adds a series of useful
information for SDSS DR7 galaxies, in particular, 
morphological classifications made by volunteers\rq{} votes. 
Whenever \lq{}ellipticals\rq{} or `spirals\rq{} are seen in our 
following analyses, the classification is based on GZ2. Out of all galaxies 
cross-matched between SDSS DR7 and GZ2, 97,785 are ellipticals 
and 135,634 are spirals. 

\subsection{SDSS groups}

Given that galaxy groups are defined as galaxies that reside in the
same dark matter halo, galaxy groups can be used to directly
probe the connections between galaxies and their host halos.
\citet{yang05, yang07} have developed a halo-based group finder 
optimized for grouping galaxies in common dark matter halos. The performance
of this group finder has been tested extensively using mock
galaxy redshift surveys constructed from CLF models 
\citep{yang03, vandenbosch03, yang04} and from a semi-analytical 
model \citep{kang05}. It was found that this group finder
is more successful than the traditional friends-of-friends (FoF)
algorithm in grouping galaxies into their common dark matter haloes 
\citep[see][(Y07 hereafter)]{yang07}. The group finder performs
consistently even for very poor systems such as isolated
galaxies in small mass haloes, which enables its suitability
to probe the galaxy-halo connection over a wide range of 
different haloes.

In the present paper, we use the DR7 group catalog, publicly 
available at \url{http://gax.shao.ac.cn/data/Group.html} to associate
galaxies with groups. This catalog is made basically by applying
exactly the same group finder of Y07  to SDSS DR7 galaxies. The
details of the group finder is described in Y07. \textit{WMAP}5 
cosmology was used to calculate distances from redshifts
and to assign halo masses to selected groups.
We adopt the group catalog \lq{}modelC\rq{}, which
uses model magnitudes rather than Petrosian magnitudes.
For each group in the group catalog, the fraction, $f_{edge}$,
of each group\rq{}s volume that falls inside of the SDSS DR7
survey volume is given. Only groups with $f_{edge}\geq0.6$
are used here, which removes about $1.6\%$ of all groups.

The group halo masses $M_{\rmn{h}}$ in the catalog are estimated
using the ranking of groups either in the combined
luminosity ($L_{19.5}$) or in the combined stellar mass
($M_{\rmn{stellar}}$) of all member galaxies with
$^{0.1}M_r-5\log h\leq-19.5$. The conversion from
$L_{19.5}$ or $M_{\rmn{stellar}}$ to $M_{\rmn{h}}$ is made by
adopting the halo mass function of \citet{tinker08} and 
the method of abundance matching assuming one-to-one 
correspondence between $L_{19.5}$ ($M_{\rmn{stellar}}$) 
and $M_{\rmn{h}}$. As shown in Y07, while both $L_{19.5}$ 
and $M_{\rmn{stellar}}$ are tightly correlated with $M_{\rmn{h}}$, the 
$M_{\rmn{h}}$ - $M_{\rmn{stellar}}$ relation is slightly tighter, 
with a typical dispersion of $\approx 0.2$ dex in $M_{\rm h}$ 
for a given $M_{\rm stellar}$ over the halo mass range considered here.
We therefore use $M_{\rmn{h}}$ based on $M_{\rmn{stellar}}$, 
although our tests showed that using $L_{19.5}$  does not 
change any of our results. For very small groups, no masses 
are assigned, and  they are excluded from our analysis.

The identification of central galaxies for each group is also
provided in two different ways: the brightest galaxy or the
most massive galaxy in terms of stellar mass. In this study, we
choose the latter as the definition of centrals. 
 
\section{An Observational Proxy of Halo Assembly Time}
\label{sec_theproxy}

\begin{figure*}
\includegraphics[width=1.0\linewidth]{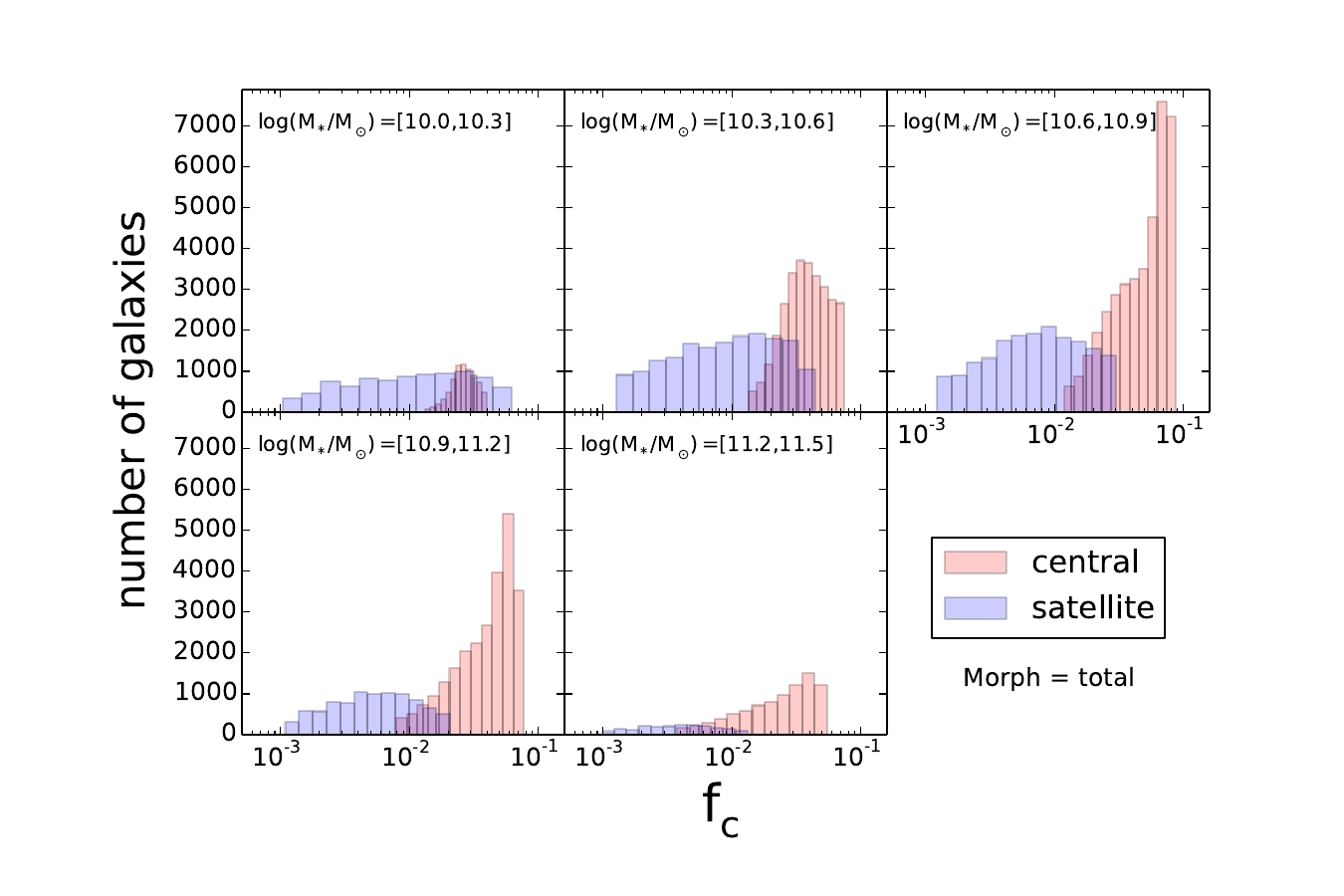}
\caption{The number distribution of galaxies in $f_c$, with each panel corresponding to different stellar mass 
bins, as indicated on the top of each panel, for centrals (red) and satellites (blue).}
\label{fig_fcdistr} 
\end{figure*}

\begin{figure*}
\hspace*{0.1cm}
\includegraphics[width=1.0\linewidth]{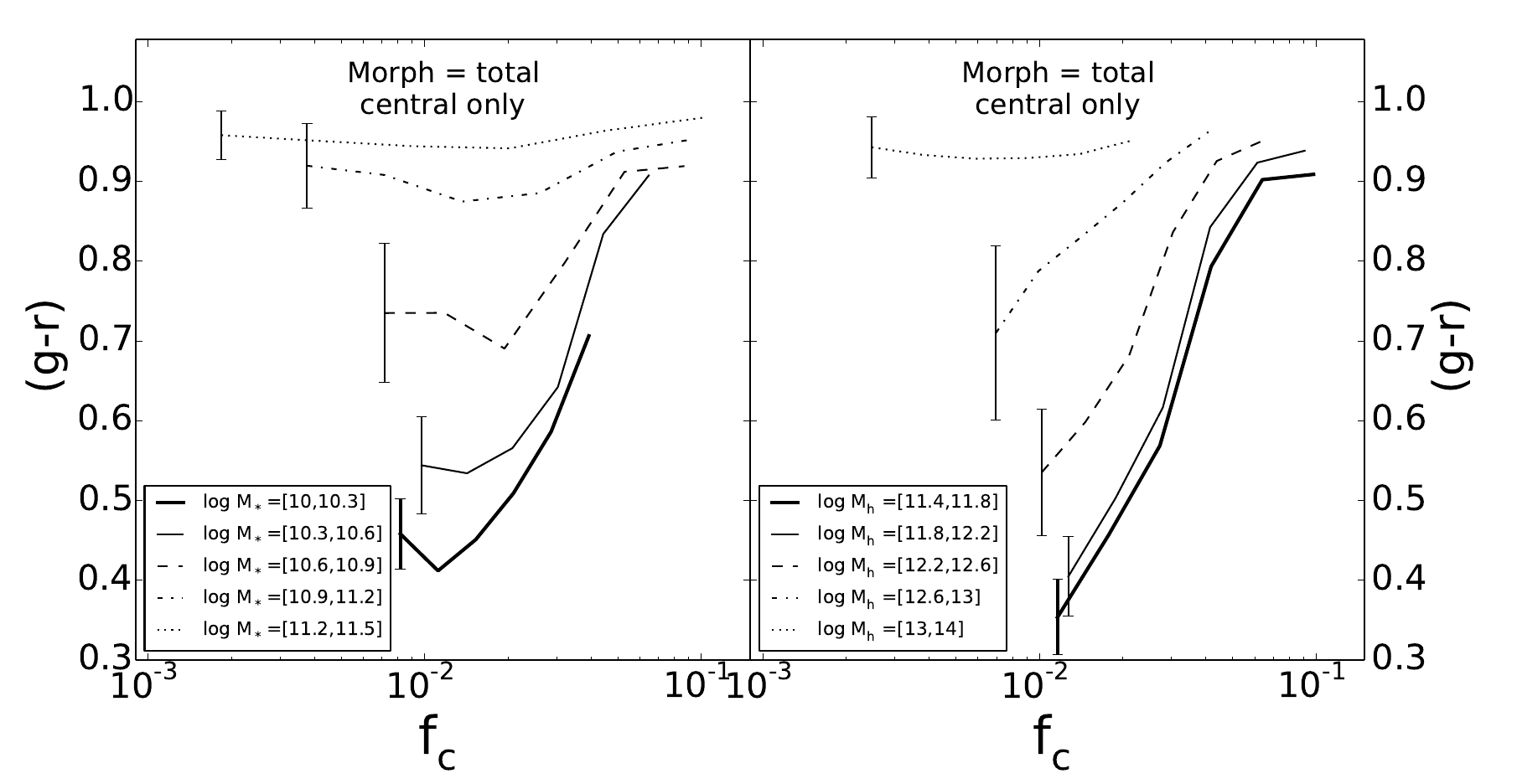}
\caption{The correlation between $(g-r)$ color, $K+E$ corrected to $z=0.1$,  and $f_c$, for centrals. 
The curves plot the median values in $f_c$ bins. 
The error bars on the leftmost  sides are `typical' $[16\%, 84\%]$ ranges for each mass bin.
In the left panel different curves refer to galaxies in different stellar mass bins, as indicated,
while in the right panel different curves are for galaxies residing in halos in 
different halo mass bins.
}
\label{fig_Ccolor} 
\end{figure*}

As mentioned in the introduction, \citet{wang11} (hereafter W11)
explored the correlations among various halo properties using 
dark matter halos identified from high-resolution $N$-body simulations.  
One of the most important properties of a halo is its formation time, $z_f$, 
which is defined to be the redshift at which the main progenitor of 
the halo  has first assembled half of its final mass. 
This formation time is believed to have 
significant impact on the properties of the galaxies the halo hosts, 
such as galaxy age, color, star formation rate (SFR), etc. Unfortunately,
$z_f$ itself is not directly observable, and so it is not 
possible to examine the correlation between $z_f$ of a halo
and the properties of the galaxies the halo host. However,     
as shown in figure 1 of W11, and reproduced in the left panel 
of Figure \ref{fig_proxy},  the halo formation time $z_f$ 
shows a tight correlation with the sub-structure fraction,
$f_s=1-f_{\rm main}$ with $f_{\rm main}=(M_{\rm main}/M_{\rm h})$,
where $M_{\rmn{main}}$ is the mass of the main sub-halo 
at the center of each host halo, quite independent of the mass
of the host halo. This suggests that $f_{\rm main}$ can be used as a proxy of $z_f$.  
Since $M_{\rmn{h}}$ can be estimated for halos using halo 
abundance matching, as described in the previous section, 
and $M_{\rm main}$ can be estimated using sub-halo
abundance matching, we can define an `observable' quantity,      
\begin{equation}
f_c\equiv {M_{*,c}\over M_{\rm h}}\,, 
\end{equation}
as a proxy of $z_f$. Here $M_{*,c}$ is the stellar 
mass of the central galaxy obtained from the rank of $M_{\rmn{main}}$. 
If there were no scatter in the halo-galaxy abundance matching,
so that there is a one-to-one relation between galaxy stellar 
mass and sub-halo mass, $M_{*,c}$ would be 
a perfectly faithful indicator of $M_{\rmn{main}}$.
By definition $M_{*,c}$ would also be the stellar 
mass of the most massive galaxy in a group because the main 
sub-halo is the most massive one among all sub-halos. 
In reality, however, the halo mass - galaxy mass relation
may not be one-to-one. Given this and that $f_s$ is not perfectly 
correlated with $z_f$, $f_c$ defined above can only be used 
as a proxy of $z_f$. As an illustration, the middle and
right panels of Fig. \ref{fig_proxy} shows the correlation 
between $z_f$ and $f_c$ obtained from the HAM model 
of \citet{hearin13} applied to dark matter halos in a high-resolution 
$N$-body simulation. As one can see, there is a tight correlation 
between $f_c$ and $z_f$ both for halos of a given mass (middle 
panel) and for centrals of a given stellar mass (right panel).
In particular, the $z_f$ - $f_c$ relation does not seem to depend 
strongly on halo mass or on galaxy mass, although massive 
systems extend further towards the low-$f_c$ end
because of the fact that $M_{*.c}$ only increases slowly with 
halo mass at the massive end \citep[e.g.][]{yang12}. 
All these validate the use of $f_c$ as an observational proxy of $z_f$. 

In what follows we will examine how galaxy properties are
correlated with $f_c$, and use the results to understand 
the connection between galaxy properties and halo assembly 
histories as represented by the formation redshift $z_f$. 
For reference, we show the distribution of galaxies in $f_c$ 
for the entire SDSS DR7 sample in Figure \ref{fig_fcdistr}. Each panel 
corresponds to a given stellar mass bin, as indicated in individual
panels, and results are shown for both centrals and satellites.  
As expected, the centrals, defined to be the most massive ones
in groups, have on average a higher $f_c$ value than satellites, 
since groups with lower $f_c$ tend to have more satellites in them.

\section{Correlation of galaxy properties with \boldmath{$\lowercase{f_{\rm c}}$}}
\label{sec_correlation}

This section examines the correlation of galaxy intrinsic 
properties with the value of $f_{\rm c}$ of the host group in which 
the galaxy resides. Results will be shown separately for central  
and satellite.  While our presentation includes all groups, 
our test using only groups with more than one member galaxies 
brighter than $^{0.1}M_r-5\log h=-19.5$ gives qualitatively 
similar results. 

\subsection{Central galaxies}

\subsubsection{Color and star formation}

\begin{figure}
\includegraphics[width=0.93\linewidth]{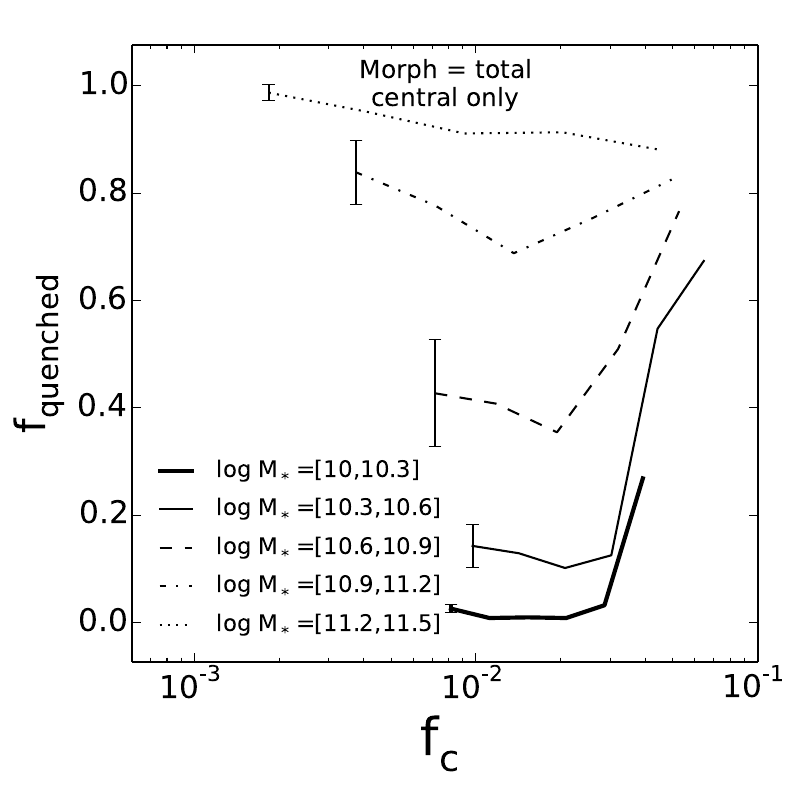}
\caption{The correlation between the fraction of quenched galaxies and $f_c$ for central galaxies. 
Quenched galaxies are defined to be the ones with star formation rate lower than the 
devision line defined by equation (\ref{eq_quench}).    
The curves plot the quenched fractions in $f_c$ bins. 
Different curves refer to galaxies in different stellar mass bins, as indicated.  
The error bars here are `typical' $1$-$\sigma$ dispersions among $100$ bootstrap 
re-sampling.} 
\label{fig_Cquench} 
\end{figure}

\begin{figure}
\vspace*{-0.2cm}
\includegraphics[width=0.99\linewidth]{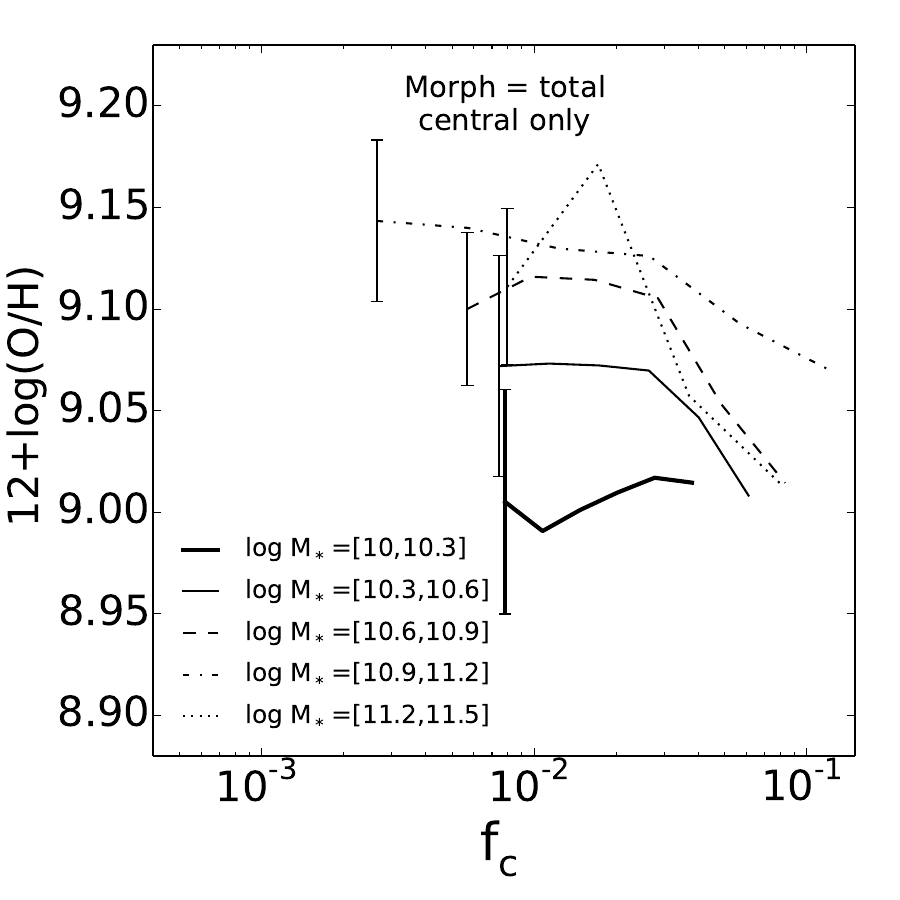}
\caption{The correlation between gas phase oxygen abundance, $12+\log\rmn{(O/H)}$, 
and $f_c$ for central galaxies. 
The curves plot the median values in $f_c$ bins. 
The error bars on the leftmost  sides are `typical' $[16\%, 84\%]$ ranges for each mass bin.
Different curves refer to galaxies in different stellar mass bins, as indicated.
} 
\label{fig_Cmetal} 
\end{figure}

\begin{figure}
\hspace*{0.1cm}
\includegraphics[width=0.98\linewidth]{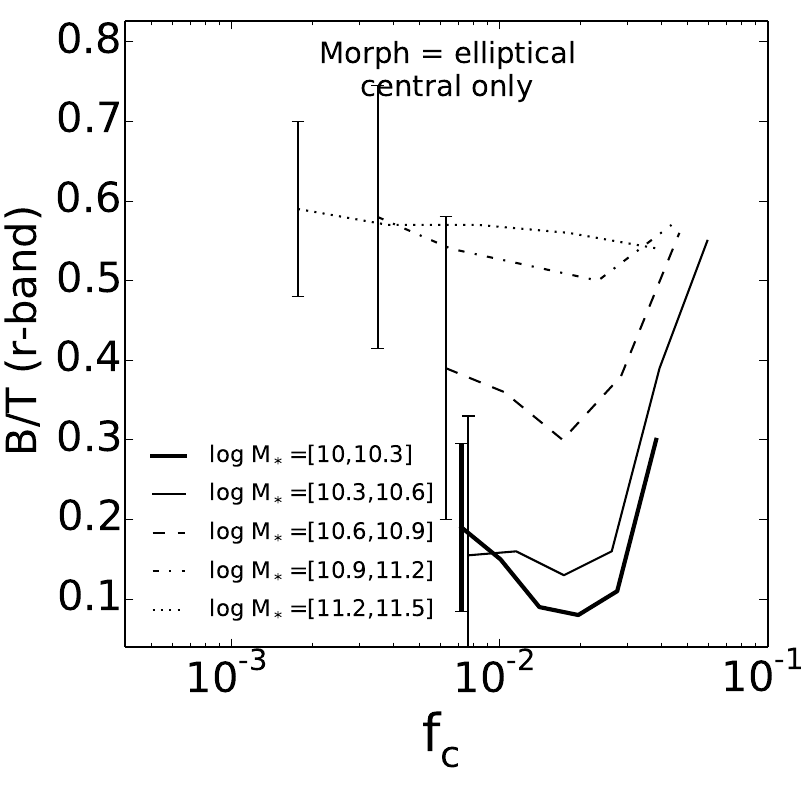}
\caption{The correlation between the bulge to total ratio (in {\it r}-band), $B/T$, 
and $f_c$ for central galaxies. 
The curves plot the median values in $f_c$ bins. 
The error bars on the leftmost  sides are `typical' $[16\%, 84\%]$ ranges for each mass bin.
Different curves refer to galaxies in different stellar mass bins, as indicated. 
} 
\label{fig_CBtoT} 
\end{figure}

\begin{figure}
\includegraphics[width=0.95\linewidth]{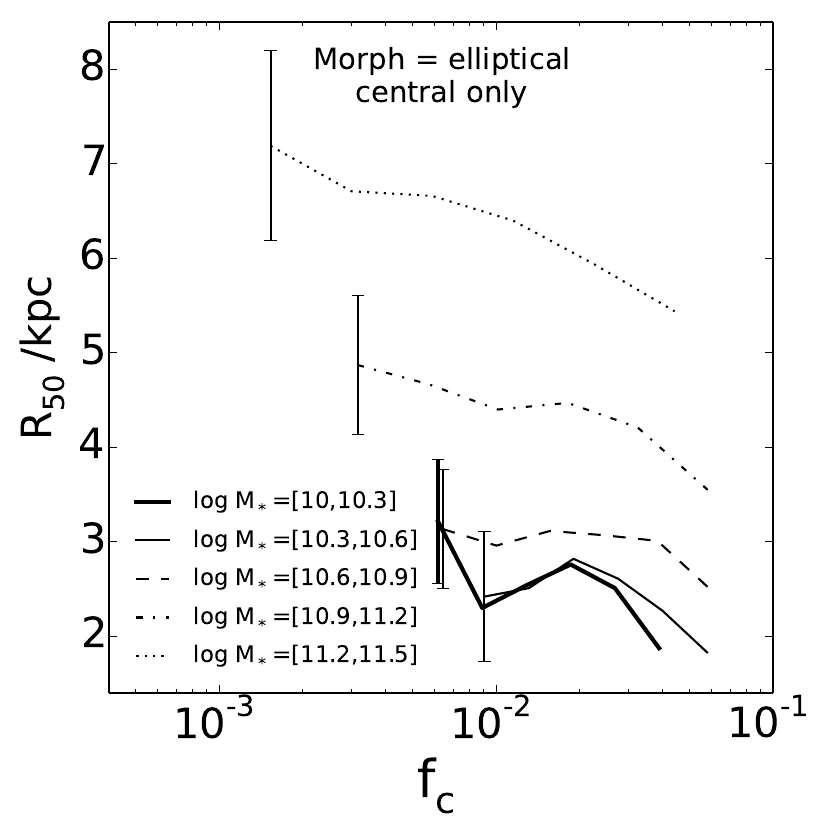}
\caption{
The correlation between the half-light radius ($r$-band) $R_{50}$ 
and $f_c$ for central ellipticals. 
The curves plot the median values in $f_c$ bins. 
The error bars on the leftmost  sides are `typical' $[16\%, 84\%]$ ranges for each mass bin.
Different curves refer to galaxies in different stellar mass bins, as indicated. 
} 
\label{fig_CsizeE} 
\end{figure}

\begin{figure}
\hspace*{0.07cm}
\includegraphics[width=0.93\linewidth]{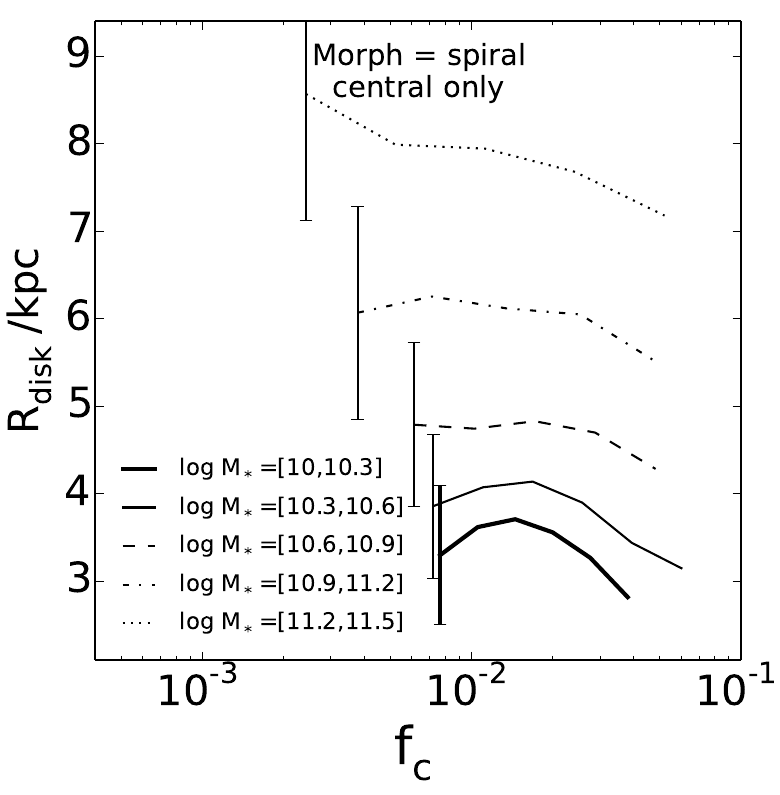}
\caption{
The correlation between disk scale length ($r$-band) $R_{\rm disk}$ 
and $f_c$ for central spirals. 
The curves plot the median values in $f_c$ bins. 
The error bars on the leftmost  sides are `typical' $[16\%, 84\%]$ ranges for each mass bin.
Different curves refer to galaxies in different stellar mass bins, as indicated. 
}
\label{fig_CsizeS} 
\end{figure}

Figure \ref{fig_Ccolor} shows the correlation between 
the $(g-r)$ color of central galaxies and $f_c$ of their host groups.
In the left panel, results are shown separately for galaxies in five different 
stellar mass ranges, as indicated in the inner panel,
while in the right panel results are shown separately 
for five different halo mass bins. The lines are the median values within narrow 
$f_c$ bins, while the bars present the typical $[16\%, 84\%]$ 
range of the distribution in the corresponding halo mass or stellar mass range.  

As one can see from the right panel, for a given halo mass, the $(g-r)$ color 
depends strongly on $f_c$, with centrals in halos with higher $f_c$ being 
redder,  except for the most massive halos, where the centrals are all equally red. 
Note that for halos with masses below $10^{12.6} {\rm M}_\odot$, the 
dependence of color on halo mass is not strong for a given $f_c$.   
By definition, for a given halo mass,  $f_c$ is directly proportional to $M_{*,c}$,  
and it is well known that the intrinsic properties of galaxies depend 
strongly on their stellar mass. Thus, the strong dependence of color on $f_c$
for a given halo mass bin see here is not surprising. However, given 
that $f_c$ is strongly correlated with halo formation time (see Fig.\,\ref{fig_proxy}), 
our results suggest that halo formation time may play an important role in determining 
the color of the central galaxies. This is demonstrated more clearly in the left panel
of  Figure \ref{fig_Ccolor}, where the $(g-r)$ color is shown as a function of $f_c$ 
for centrals of fixed stellar mass.  As one can see, massive galaxies are more or less all red, 
independent of $f_c$, while for galaxies with $M_*<10^{11}{\rm M}_\odot$, 
their colors depend strongly on $f_c$. There seems to be a characteristic 
value $f_c\sim  0.01$ - $0.02$,  below and above which the color shows the 
opposite trends with $f_c$. At the high $f_c$ end, galaxies become
increasingly redder as $f_c$ increases,  which may be produced by
the fact that groups with higher $f_c$ on average assembled their halos earlier.    
In contrast, galaxies in groups with $f_c<0.02$ seem to have 
a reversed, albeit weak, trend between color and $f_c$.
Note that for a given central stellar mass, lower $f_c$ corresponds 
to higher halo mass.  The reversed trend at low $f_c$ 
reflects a `down-sizing' effect of massive halos, in that centrals
in massive halos formed their stars earlier than in low mass halos 
\citep[e.g.][]{Lu_etal2015}, although the massive host halos themselves  
assembled (half of their masses) later (lower $f_c$).
This is consistent with the fact that {\it in situ} star formation 
in massive halos is quenched once their masses reached
a few times $10^{12} {\rm M}_\odot$ \citep[e.g. figure 14 of][]{Lu_etal2014}, 
and, for high mass halos, more massive ones on average 
assembled a fixed amount of mass earlier \citep{LiMoGao2008}.    

Figure \ref{fig_Cquench} shows the quenched fraction of centrals
as a function of $f_c$. Because for a given halo mass,  
$f_c$ and stellar mass is strongly degenerated for centrals, 
here and in the following we only show results for centrals divided 
into different stellar mass bins but not divided further 
according to halo mass. For galaxies in each stellar mass bin, we separate them 
into quenched and star forming sub-populations using the definition of 
\citet{moustakas13}, 
\begin{eqnarray}\label{eq_quench}
\log\left(\frac{\rm SFR}{\rm M_\odot yr^{-1}}\right)&=&
-0.49+0.65\log\left(\frac{M_*}{10^{10}\rm{M_\odot}}\right)\nonumber\\
&&+1.07(z-0.1)\,. 
\end{eqnarray}
For a given $M_*$, galaxies with star formation rate (SFR) above the 
value given by the above equation are defined to be star forming, and 
those with SFR below the value are defined to be quenched.  Given 
that the the specific star formation rate (sSFR, 
defined as the ratio between SFR and $M_*$) of 
a galaxy is closely related to its color, it is not surprising that the general 
trends seen in this plot are similar to those shown in Fig.\,{\ref{fig_Ccolor}. 
Low-mass centrals are dominated by star forming 
galaxies in halos of low $f_c$ but become dominated by quenched 
galaxies at the high end of $f_c$. A reversal of trend is again 
seen at $f_c\sim 0.02$. 
   
Finally, let us look at the gas phase metallicity of galaxies, which is 
shown as a function of $f_c$ in Figure \ref{fig_Cmetal}. The 
gas phase metallicity estimates are available only for a limited fraction 
of galaxies, mostly star forming ones. The result for the highest stellar 
mass bin is quite noisy because here only a small fraction of galaxies 
are star forming.  For a given stellar mass, 
there is a clear trend that the gas phase metallicity decreases with 
increasing $f_c$. For centrals with $M_{*}> 10^{10.3}{\rmn{M}}_\odot$,
the decrease with $f_c$ is quite rapid, by almost 0.1 dex. 
This decrease is comparable to the scatter in the gas phase metallicity - stellar 
mass relation obtained by \citet{tremonti04}, suggesting that 
the scatter may be dominated by the variance in halo assembly, 
with galaxies formed in older dark matter halos tend to have lower 
gas-phase metallicities.  

\subsubsection{Structure and size}

The bulge to total ratio, $B/T$, as described in the data section, 
is plotted against $f_c$ in Figure \ref{fig_CBtoT}.
 There are a number of interesting trends. Overall, the $B/T$ increases 
with stellar mass, simply owing to the fact that earlier type galaxies 
are on average more massive. For massive galaxies with 
$M_*$ higher than about $10^{11}{\rm M}_\odot$, the 
$B/T$ ratio on average decreases with  $f_c$. For galaxies 
with lower stellar masses, the trend changes at $f_c\sim 0.02$. 
While the $B/T$ ratio decreases with increasing $f_c$ at 
the low $f_c$ end, it increases with $f_c$ rapidly at $f_c>0.02$.  

As mentioned above, for a given central mass $M_{\rm *,c}$, 
a lower $f_c$ on average corresponds to a higher halo mass $M_{\rm h}$. 
Since a higher halo mass  on average corresponds to a higher 
group richness, the decline of $B/T$ with $f_c$ may, therefore,
be understood in terms of the  morphology-density relation found 
by \citet{Dressler1980} that early-type galaxies 
(higher $B/T$) are preferentially found in high density environments, 
while late-type galaxies are more likely to be found in poor groups and 
in the lower density fields.  The increase of $B/T$ with 
decreasing $f_c$ at the low $f_c$ end shown in 
Fig.\,\ref{fig_CBtoT} follows such a morphology-density relation. 
However, our results also contain new information, in that the 
morphology-density relation is present even for centrals of a given 
stellar mass.  

\begin{figure*}
\includegraphics[width=0.8\linewidth]{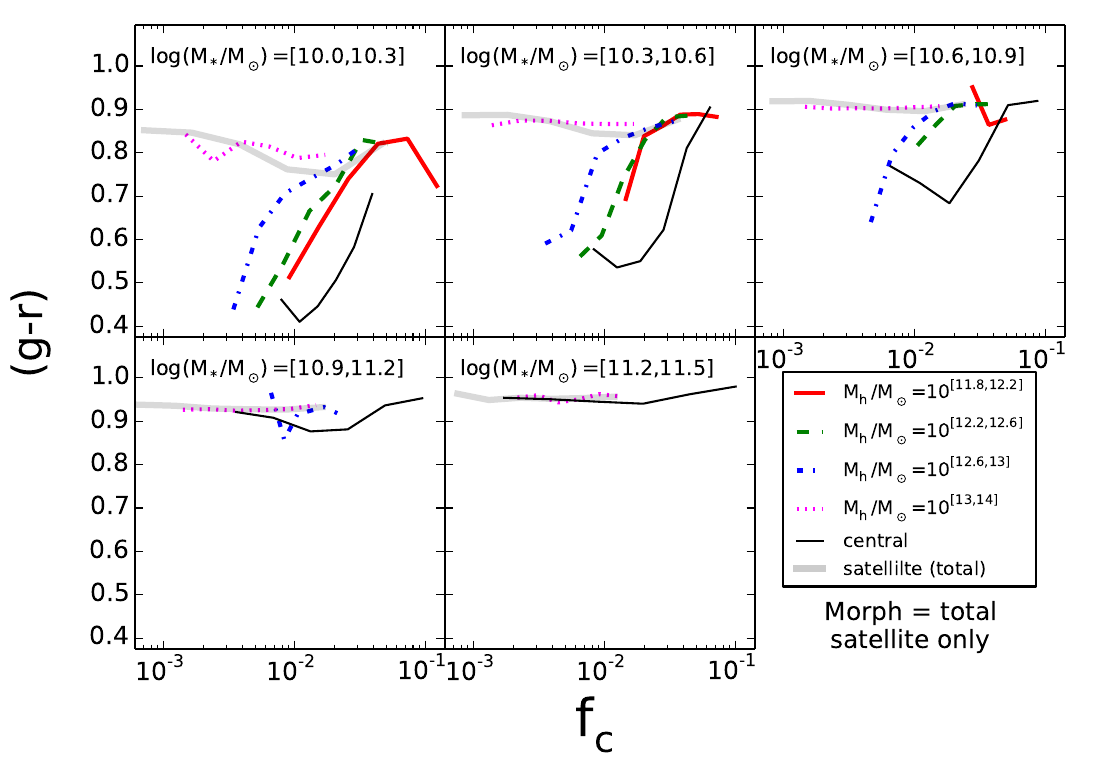}
\caption{
The correlation between $(g-r)$ color and $f_c$ for satellite galaxies (thick lines). 
Individual panels show the medians in $f_c$ bins for satellites of different stellar masses.
Within each panel, satellites are divided into  four subsamples
according to the masses of their host halos, as denoted in the legend. 
The result for the total satellite sample in a given stellar mass bin 
is shown as the translucent thicker line in each panel. For comparison   
medians for centrals shown in Fig.\,\ref{fig_Ccolor}
are re-plotted here as the thin solid lines.
} 
\label{fig_Scolor} 
\end{figure*}

\begin{figure*}
\includegraphics[width=0.8\linewidth]{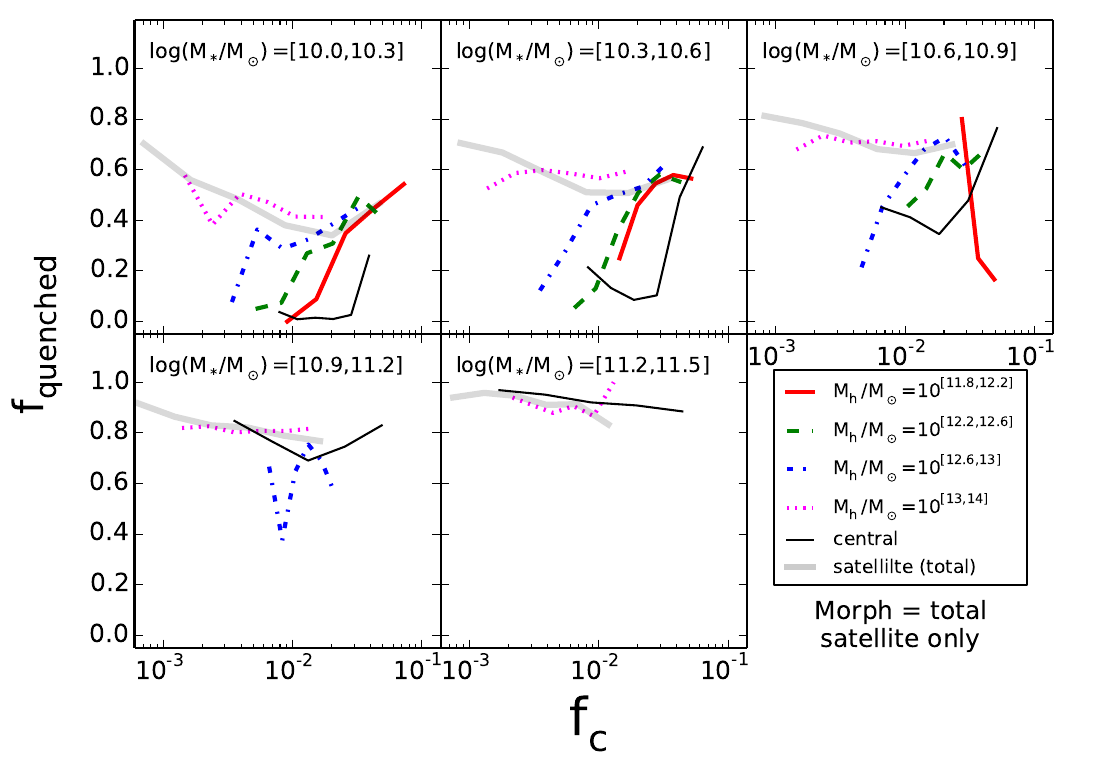}
\caption{
The correlation between the quenched fraction and $f_c$ for satellite galaxies (thick lines). 
Individual panels show the medians in $f_c$ bins for satellites of different stellar masses.
Within each panel, satellites are divided into  four subsamples
according to the masses of their host halos, as denoted in the legend. 
The result for the total satellite sample in a given stellar mass bin 
is shown as the translucent thicker line in each panel. For comparison   
medians for centrals shown in Fig.\,\ref{fig_Cquench}
are re-plotted here as the thin solid lines.
} 
\label{fig_Squench} 
\end{figure*}

The strong increase of $B/T$ with increasing $f_c$ seen 
for low-mass central galaxies with $M_\star <10^{11}{\rm M}_\odot$
at $f_c>0.02$ runs against the morphology-density relation. 
Since larger $f_c$ means an earlier assembly time, as 
shown in the last section,  the trend of $B/T$ with $f_c$
indicates an dependence on halo assembly time, in that 
central galaxies in older halos tend to have higher $B/T$. In the
current CDM paradigm of structure formation, the formation of 
halos of a given mass at earlier time is on average more 
dominated by major mergers and older halos are 
on average  more compact \citep[e.g.][]{Li_etal2007,Zhao_etal09}.
If the bulge components are formed through 
major mergers or through secular evolutions of the disk 
components, their formation is expected to be promoted by 
both major mergers and a compact structure of dark matter halos.
The positive correlation between $B/T$ and $f_c$ obtained here 
may follow directly from such formation. The reversal of the trend   
at $f_c<0.02$ is also consistent with such interpretation, because 
central galaxies in massive halos actually have earlier formation
due to the down-sizing effect described above. 
  
Figures \ref{fig_CsizeE} and \ref{fig_CsizeS} show how the sizes
of central galaxies of a given stellar mass correlate with $f_c$.
Results are shown separately for the half-light radius 
($R_{50}$) of ellipticals and the disk scale-length 
($R_{\rm disk}$) of spiral galaxies. Here the morphological 
separation is made according to the visual classification from 
GZ2, and the sizes are taken from the $r$-band bulge-disk 
decompositions of \citet{simard11}.  For both ellipticals and spirals, 
more massive galaxies are larger, as expected. For a given stellar 
mass, the sizes of centrals decrease with $f_c$ at $f_c>0.02$. 
This is consistent with the interpretation that  halos formed earlier 
on average are smaller. However, unlike star formation, there is 
no strong reversal of trend at $f_c<0.02$, in particular for
massive galaxies. For elliptical galaxies,  this may be due 
to the fact that the assembly of the stellar component follows 
halo assembly more closely than star formation 
\citep[e.g. figure 5 in][]{Lu_etal2015}. For spiral galaxies, this 
result may indicate that disks can continue to accrete cold gas
from halos as the halos grow, even in relatively massive systems.  

\subsection{Satellite galaxies}

\begin{figure*}
\includegraphics[width=0.8\linewidth]{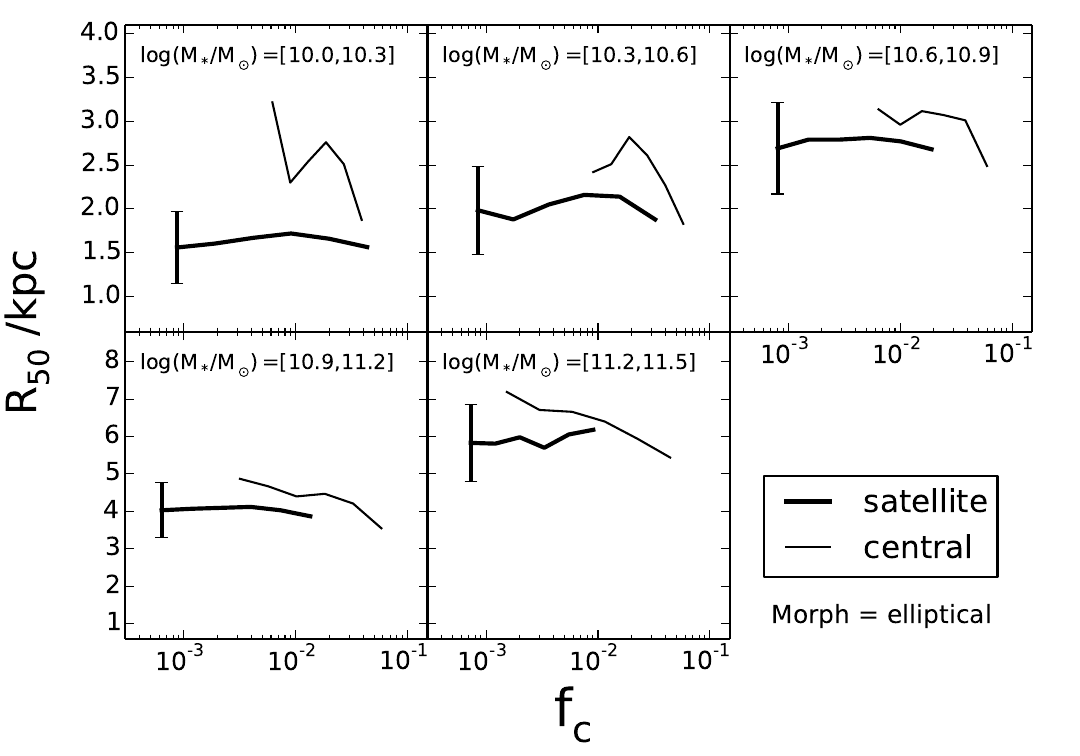}
\caption{The correlation between $R_{50}$ and  $f_c$ for satellite ellipticals.
The thick curves are the medians in $f_c$ bins, while  the `typical' 
$[16\%,84\%]$ ranges are indicated by the bars on the leftmost sides. 
For comparison,  results for central ellipticals shown in
Fig.\,\ref{fig_CsizeE} are re-plotted here as thin lines. 
Different panels show the results  in different stellar mass bins,
as indicated. } 
\label{fig_SsizeE} 
\end{figure*}

\begin{figure*}
\includegraphics[width=0.8\linewidth]{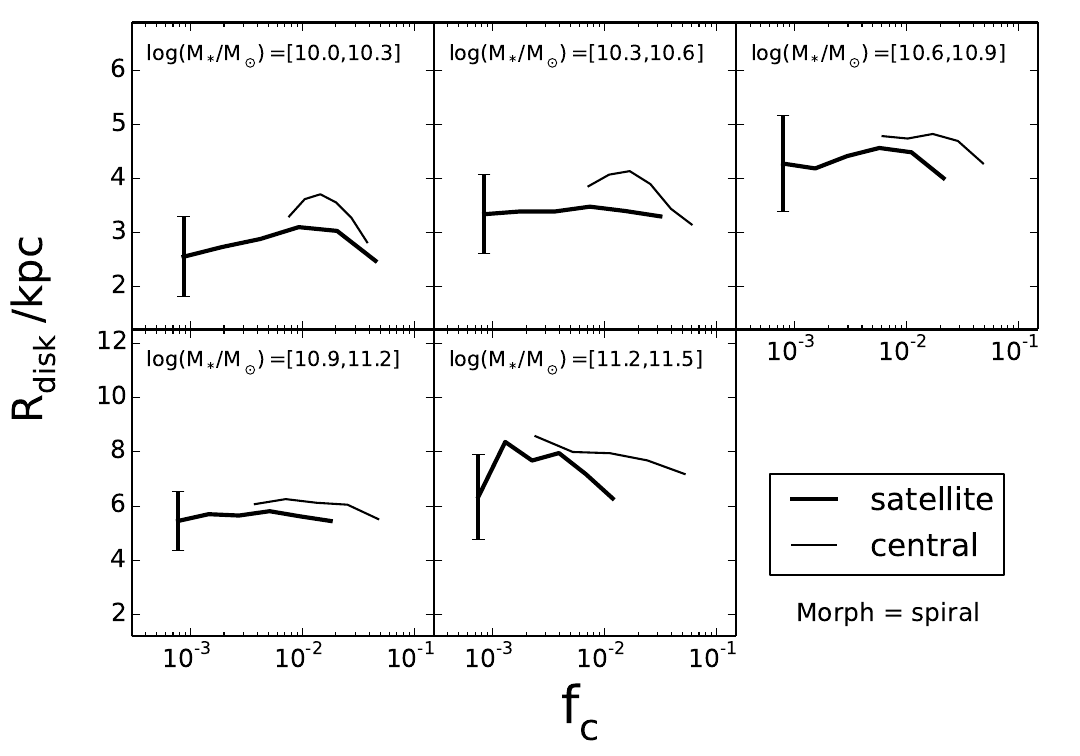}
\caption{The correlation between $R_{\rm disk}$ and  $f_c$ for satellite spirals.
The thick curves are the medians in $f_c$ bins, while  the `typical' 
$[16\%,84\%]$ ranges are indicated by the bars on the leftmost sides. 
For comparison,  results for central spirals shown in
Fig.\,\ref{fig_CsizeS} are re-plotted here as thin lines. 
Different panels show the results  in different stellar mass bins,
as indicated. } 
\label{fig_SsizeS} 
\end{figure*}

  Figure \ref{fig_Scolor} shows the $(g-r)$ color of satellite 
galaxies  as a function of $f_c$ of their host groups. The 
five panels show the results of galaxies in five stellar mass 
bins, as indicated. For each stellar mass bin, results are shown 
separately for galaxies in groups of four different halo mass 
bins, as indicated in the small panel.  
For given $f_c$ and halo mass, more massive galaxies 
on average are redder. For the most massive galaxies with 
$M_*>10^{11}{\rm M}_\odot$,  which are only found in massive halos, 
their $(g-r)$ colors are all red, quite independent of $f_c$. For satellites 
with lower stellar masses ($M_*<10^{11}{\rm M}_\odot$), 
there is a marked trend that the $(g-r)$ color becomes increasingly 
redder as $f_c$ increases. The trend is weaker for groups with 
higher halo masses, and becomes almost totally flat for halo masses above 
$\sim 10^{13} {\rm M}_\odot$ (the magenta dotted curve in each panel). 
We do not see a reversal in the trend in any ranges of $f_c$ as seen 
in central galaxies shown in Fig.\,{\ref{fig_Ccolor} (reproduced 
here as the black solid curves for comparison),  
because here results are shown separately for groups in different 
halo mass bins. If we consider all satellites of a given stellar mass
regardless of their host halo mass, then we get the results as shown 
by the thick shaded line in each panel.  Here we do see a 
change of trend at $f_c\sim 0.02$, which is similar to, albeit weaker 
than that for central galaxies.  Clearly, satellites at the low-$f_c$ 
end are dominated by the ones in massive groups. 
The reversed trend at $f_c<0.02$, is consistent with the 
fact that galaxies in massive halos actually have earlier 
formation due to the down-sizing effect described earlier. 

 Figure \ref{fig_Squench} shows the quenched fraction 
of satellites  as a function of $f_c$ of their host groups. 
The format of this figure is exactly the same as Fig.\,{\ref{fig_Scolor}, 
and the quenched fraction is again determined by using  
equation (\ref{eq_quench}). The trends shown here are very 
similar to those in Fig.\,{\ref{fig_Scolor}, again because 
the $(g-r)$ color is closely correlated with the sSFR used to 
separate quenched from star-forming galaxies.
 
Finally let us look at the sizes of galaxies. Here we consider 
ellipticals and spirals separately. Our tests showed that the 
dependence of size on halo mass is weak for satellites 
and the current samples are too small to give significant results 
for the halo-mass dependence. Thus, we only divide galaxies into 
stellar mass bins, but not further into halo mass bins. Figures 
\ref{fig_SsizeE} and \ref{fig_SsizeS} 
show how the sizes of satellite galaxies of a given stellar mass 
correlate with $f_c$ (thick solid curves).
Here results are shown separately for 
the half-light radius ($R_{50}$) of ellipticals and the disk scale-length 
($R_{\rm disk}$) of spiral galaxies, both taken from the $r$-band 
bulge-disk decompositions of \citet{simard11}.  For both ellipticals and spirals, 
the trend with $f_c$ is rather weak, although for low-mass 
galaxies the size seems to decrease as one moves away from 
$f_c\sim 0.01$ toward both the low and high ends of 
$f_c$. This trend suggests that galaxies of a given stellar 
mass on average have smaller sizes if formed earlier.   

Compared with central galaxies of the same stellar mass (shown 
by the thin curves), satellites are smaller. This is true for both 
spirals and ellipticals, and the difference is larger for lower mass 
galaxies. It is interesting to note that the average sizes of satellites 
are comparable to those of centrals with the highest
$f_c$, which indicates that sub-halos which host satellites may have 
as early formation as the oldest halos of similar masses that host centrals. 

\citet{weinmann09} found that, at fixed stellar mass, late-type 
satellite galaxies have smaller radii than late-type central galaxies. 
Our results confirm theirs. However, \citet{weinmann09} found 
no difference in size for early-type galaxies, while 
Fig.\,\ref{fig_SsizeE} shows clearly that such difference also exists
for ellipticals, particularly for ellipticals with low stellar masses. 
The discrepancy may arise from the difference in the separation 
of early versus late types. While \citet{weinmann09} used the 
concentration parameter, defined as the ratio between $R_{90}$
(radius within which $90\%$ of the total light is included) and $R_{50}$, 
we use morphological classifications from GZ2. 
\citet{weinmann09} interpreted their finding as owing to the fading 
of stellar disks due to the aging of stars.  However, it is unclear
if such an interpretation can also explain the systematic change 
of disk size of central galaxies with $f_c$. Passive stellar evolution
alone is also difficult to explain the difference between centrals and 
satellites for elliptical galaxies. Based on our results, the more 
likely reason is that halos formed earlier are more compact, and 
that the difference in sizes between centrals and satellites is due 
to differences in formation time, just as centrals in halos of different $f_c$.  

\section {Comparison with Models}
\label{sec_theory}

\begin{figure*}
\includegraphics[width=0.95\linewidth]{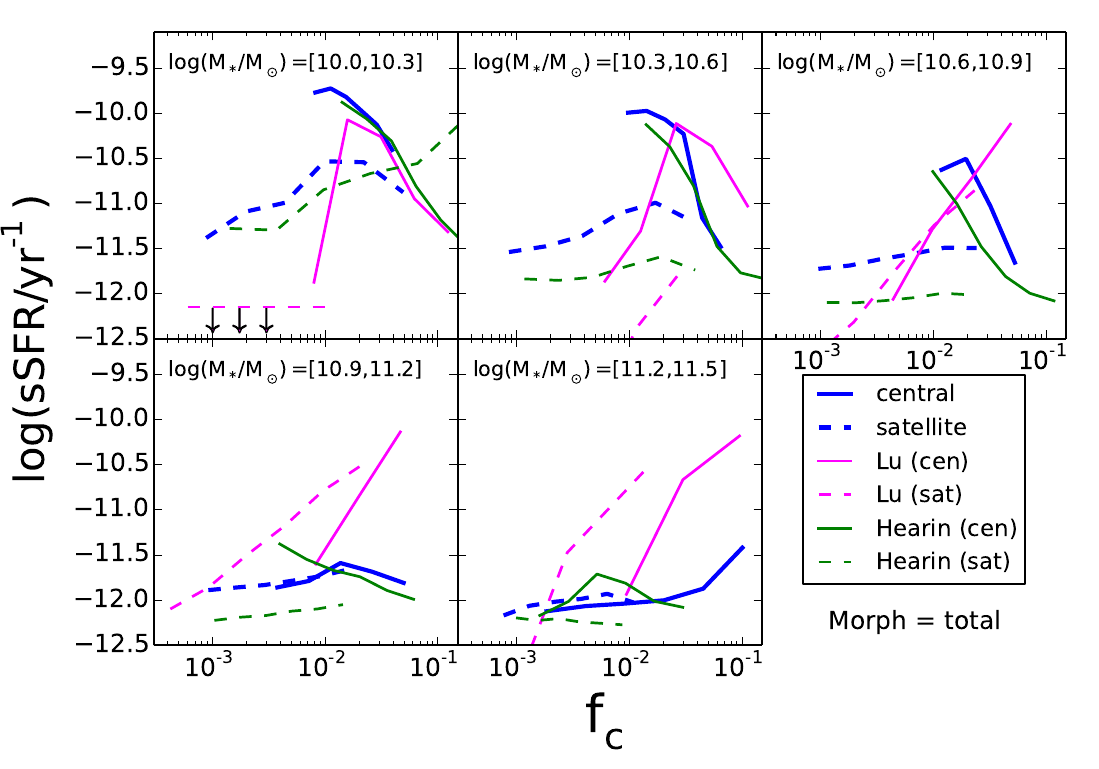}
\caption{The observed correlation between $f_c$ and the specific star formation rate, 
sSFRs (blue lines: solid for centrals and dashed for satellites) in comparison  
to the predictions of the semi-analytical model (SAM) of \citet{LuY_etal2014}
(thin magenta) and the age abundance matching model of \citet{hearin13} (thin green).
Note that the sSFR of satellites in the first panel for the SAM are too low to show, 
and are represented by a horizontal line with down pointing arrows. 
}
\label{fig_model} 
\end{figure*}

 In order to explore the implications of our findings, we make comparisons 
of our results with some theoretical models. Since our results 
are derived from galaxy groups selected from a redshift catalogue, a detailed 
comparison between our observational results with theoretical models requires
the construction of theoretical mock catalogs that take into account all observational
selection effects. This is beyond the scope of this paper, and we will come back 
to this in a forthcoming paper. In this paper, we use halo occupations 
of galaxies predicted directly by models, ignoring all observational selection effects. 
As a demonstration, we use two specific models: the empirical age abundance 
matching (AAM) model published in \citet{hearin13, hearin14}, and the semi-analytical 
model (SAM) as described in \citet{LuY_etal2014}.  

While traditional abundance matching techniques only exploit the correlation 
between luminosity of galaxies and mass of their host haloes to assign 
galaxies in haloes from simulations, the AAM connects 
galaxies to haloes as a function of both color and luminosity. Specifically, it
assigns stellar masses to galaxies according to the mass ranking of their 
host halos, and  assign colors to galaxies of a given stellar mass 
according to the formation time ranking of their halos. 
The SAM approach, on the other hand,  attempts to model physical processes
using simplified receipts parameterized in simple functional forms.   
A SAM generally contains a large number of free parameters.   
\citet{LuY_etal2014} used a Monte Carlo Markov Chain method 
to infer their model parameters from  observational constraints such as 
luminosity functions of galaxies at different redshifts. The Lu et al. SAM 
contains many of the same components as other SAMs. In particular 
it assumes a strong star formation feedback and an efficient gas stripping
to prevent too much star formation in dark matter halos. 

Since none of the models provides reliable predictions for the 
structural properties of galaxies, here we focus only on the star 
formation properties as represented by the specific star formation
rate (sSFR) of galaxies.   
Figure \ref{fig_model} shows the sSFR as a function of $f_c$ as predicted 
by the models of \citet{hearin13} and \citet{LuY_etal2014}. As in the 
observation, we identify the most massive galaxy in a halo to be 
the central galaxy, and use the ratio between $M_{c*,c}$ and the halo mass to 
define $f_c$. Here results are shown separately for centrals 
and satellites in five different stellar mass bins. For comparison,  
our observational results are included in each panel. As one can see, the AAM model  
reproduces the observational trends qualitatively. 
In particular, the rapid decreases of sSFR with increasing $f_c$ for central 
galaxies in the low stellar mass bins are well reproduced. 
The trends for satellite galaxies are also well produced, although 
the predicted sSFR are systematically lower than the observational
results. This discrepancy should not be taken too seriously, as
the satellite population in observational groups may be 
contaminated by centrals that on average have higher SFR than 
the satellites of the same mass. As mentioned above, 
such contaminations can only be taken into account properly by 
applying the same group finder to the mock catalog constructed from 
the AAM model. 

In contrast,  the predictions of the SAM are very different 
from the observational results. The model predicts 
too much quenching of star formation in low mass satellites, while 
the star formation rates in centrals, particularly in groups with high $f_c$,
are over-predicted by more than an order of magnitude.
The SAM also fails to catch the overall trends in the observation, 
even qualitatively. These results suggest that the halo assembly 
plays an important role in regulating star formation, and the underlying  
physical processes  are still poorly captured in the SAM considered here. 
It is clearly interesting to compare our results with other SAMs 
and simulation results, not only in sSFR, but also in other properties, 
such as size,  $B/T$, and metallicity, to explore the implications of our results.

\section{Summary}
\label{sec_summary}

We have showed that the ratio, $f_c\equiv M_{*,c}/M_{\rmn{h}}$, can be used as a reliable 
observational proxy of halo assembly time, with higher $f_c$ for halos that assembled earlier. 
This use was motivated by the results of W11, who used N-body simulations to show
that there is a tight correlation between $M_{\rm main}/M_{\rmn{h}}$ 
($M_{\rm main}$ being the main sub-halo mass) and halo half-mass 
assembly redshift ($z_f$), combined with (sub)halo abundance matching.
We used the SDSS groups by Yang et al. to investigate how galaxy properties 
are correlated with the assembly times of their host halos. 

Central galaxies of a given stellar mass with higher $f_c$ are found to be redder and 
more quenched in star formation while $f_c>0.02$. This implies that star formation 
in centrals in this regime is dictated by their halo assembly history. A
reversed albeit weak trend is seen for centrals with $f_c<0.02$, which reflects 
the down-sizing effect that a more massive halo on average reaches 
the mass of most efficient {\it in situ} star formation,  $\rmn{\sim 10^{12}M_{\odot}}$, 
earlier. Similar trends with $f_c$ are found for the bulge to total ratio, $B/T$: 
central galaxies hosted by older halos tend to have higher $B/T$ ratios. 
We suggest that this is because older halos are more compact and their 
formation is more dominated by major mergers.  For a given stellar mass, 
the sizes of central galaxies are also correlated with $f_c$ for both ellipticals 
(in terms of the half-light radius, $R_{50}$) and spirals 
(in terms of the disk scale-length, $R_{\rm disk}$), with centrals hosted by 
older halos being smaller. This trend is again consistent with the fact that
halos of a given mass are more compact at higher redshifts. 

We have also analyzed how the intrinsic properties of satellite galaxies
change with the value of $f_c$ of their host halos.  Here we found that, 
for a given stellar mass, satellites residing in older halos are redder
and more quenched, and this trend is stronger for lower mass halos. 
Satellites also appear smaller than centrals of the same mass, and this 
is true for both ellipticals and spirals. These results can again be explained 
by the fact that halos that assembled earlier are more compact.  
As for centrals, a weak down-sizing effect in the quenching of star formation 
is also seen for satellites hosted by massive halos with $f_c<0.02$.
These results, together with those found for the centrals, demonstrate
clearly that halo assembly plays an important role in determining the 
properties of galaxies the halos host.   

We present our preliminary comparisons of our observational results with
the predictions by the AAM model of \citet{hearin13} and by 
the SAM of \citet{LuY_etal2014}. The AAM model reproduces well 
the general trends in the observational data, while the SAM fails to do so. 
The SAM predicts too many quenched low-mass satellites and
too small fraction of  quenched high-mass galaxies. These 
imply that halo assembly history is another important factor in addition 
to halo mass that can affect star formation in galaxies, and such effects 
have yet to be properly modeled in the SAM. In this context, the observational 
results obtained here are expected to provide stringent   
constraints on theoretical models of galaxy formation and evolution. 
We will come back to detailed comparison between our observational 
results and model predictions in a forthcoming paper.  
 
\section*{Acknowledgments}
We thank Andrew Hearin for help in using his mock catalog, Yu Lu for using his 
semi-analytical data, and Zhankui Lu for helpful discussion and the anonymous
referee for helpful comments that greatly improved the presentation of this paper. 
HJM acknowledges the support from NSF AST-1109354. This work is also supported 
by the 973 Program (Nos. 2015CB857002,2015CB857005).


\label{lastpage}

\end{document}